# A Raspberry Pi-Based Attitude Sensor

A. G. Sreejith[1,†], Joice Mathew[1], Mayuresh Sarpotdar[1], Rekhesh Mohan[1], Akshata Nayak[1,2], Margarita Safonova[1] and Jayant Murthy[1]

[1] *Indian Institute of Astrophysics, Bangalore, India*
[2] *Jain University, Bangalore, India*
†*agsreejith@iiap.res.in*

We have developed a lightweight low-cost attitude sensor, based on a Raspberry Pi, built with readily available commercial components. It can be used in experiments where weight and power are constrained, such as in high-altitude lightweight balloon flights. This attitude sensor will be used as a major building block in a closed-loop control system with driver motors to stabilize and point cameras and telescopes for astronomical observations from a balloon-borne payload.

## 1. Introduction

We have recently begun a program to use low-cost high-altitude balloons for astronomical and atmospheric observations [Nayak et al., 2013]. The main aim is the observations of extended nearby objects (e.g. comets) and of diffuse sources (e.g. zodiacal light or airglow) with wide field of view (FOV) instruments. A key requirement for these observations is accurate and stable pointing in spite of the wind speeds of tens of knots and local turbulence known to exist at altitudes above 30 km [Chingcuanco, 1989], which cause the payload to sway and rotate [Hazen, 1985; Nigro et al., 1985]. We are developing a 3-axis stabilization system to correct for these random motions, where the first step is development of an attitude sensor to determine the pointing direction. An initial pointing accuracy within $0.5°$ is sufficient for the wide-field instruments (1–2 degrees) we plan to use. Because we are using lightweight balloons, the total payload weight is limited to less than 6 kg, placing constraints on the sensor size, weight and power. In keeping with our philosophy of developing low-cost balloon experiments, we use only easily sourced commercial components.

Different types of attitude sensors have been used in high-altitude balloon flights: star trackers, sun sensors, magnetometers, accelerometers and gyroscopes, each with their own advantages and disadvantages. We have chosen to use a commercial inertial measurement unit (IMU), which combines an accelerometer, a magnetometer and a gyroscope on a single chip. The IMU provides attitude information in terms of Euler angles and quaternions [Bar-Itzhack & Oshman, 1985], which we combine on a Raspberry Pi[1] with the output of a global positioning system (GPS) chip to obtain the celestial coordinates of the pointing direction. The construction of a lightweight compact attitude sensor has become much simpler in recent years due to the replacement of mechanical devices by solid state devices such as, for example, micro-electromechanical systems (MEMS). In this paper we describe the development of a Raspberry Pi-based attitude sensor which uses onboard data fusion, and describe our plan to expand it into a fully-functional 3-axis stabilization system with a star tracker.

## 2. Hardware and Implementation

We have built our attitude sensor around a Raspberry Pi, a small-size single-board computer (SBC) developed by the Raspberry Pi Foundation (technical specifications in Table 1). The Raspberry Pi and similar SBCs (e.g. the Arduino) provide flexible, low-cost platforms to which different devices may be connected, and which have been used in a number of innovative ways from home automation systems to aeronautical communication devices [Sabastian et al., 2012].

---

[1] http://www.raspberrypi.org

Table 1. Raspberry Pi Technical Specifications (Source: Raspberry Pi user guide, Raspberry Pi Foundation).

| | |
|---|---|
| Chip | Broadcom BCM2835 full HD multimedia applications processor |
| CPU | 700 MHz Low Power ARM176JZ-F applications processor |
| GPU | Dual Core VideoCore Multimedia Co-Processor |
| Memory | 512MB SDRAM |
| Ethernet | Onboard 10/100 Ethernet connector |
| USB 2.0 | Dual USB connector |
| Video Output | HDMI composite RCA |
| Audio Output | 3.5mm jack, HDMI |
| Onboard Storage | SD, MMC, SDIO card slot |
| Weight | 45 gms |
| Power Rating | 700 mA (3.5 W) |
| Operating System | Raspbian |
| Dimensions | 8.6cm x 5.4cm x 1.7cm |

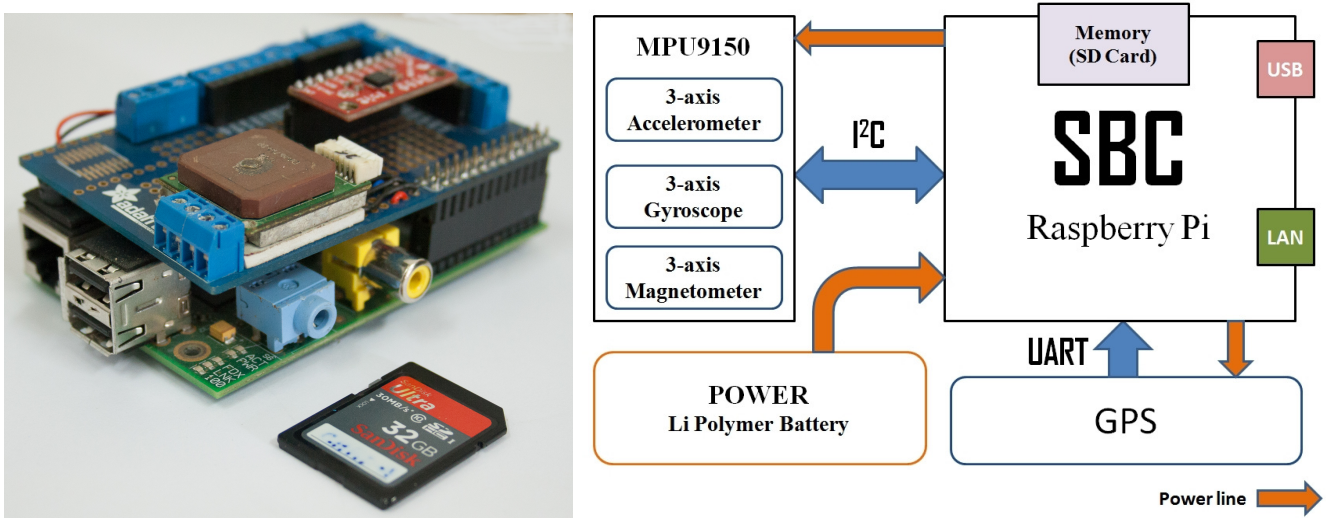

Fig. 1. Attitude sensor. *Left*: Assembled attitude sensor with an SD card for scale. *Right*: Functional block diagram of the attitude sensor.

The photograph of the sensor, along with an SD card for scale, is shown in Fig. 1, *Left*, and its functional block diagram is shown in Fig. 1, *Right*. We have used a 9-axis MEMS IMU (MPU-9150[2]) which contains two chips, the MPU-6050 comprising a 3-axis gyroscope, an accelerometer and a digital compass (Ak8975), and a Digital Motion Processor (DMP). This IMU is commonly used in a variety of commercial applications including mobile phones, tablets and gaming platforms. Gyroscopes are included because the accuracy of accelerometers and magnetometers is satisfactory only when the system is stationary and far from induced random magnetic fields [Ayub et al., 2012], whereas gyroscopes provide a fast response to changes in orientation while being independent of magnetic fields. The MPU-9150 uses the proprietary InvenSense Digital motion fusion$^{\text{TM}}$ and runtime calibration firmware to combine data from the accelerometer and the gyroscope onboard the chip.

---

[2]InvenSense Inc, USA (http://www.invensense.com/mems/gyro/mpu9150.html)

The MPU-9150 has user-programmable settings for the gyroscope with ranges of ±250, ±500, ±1000, ±1000 and ±2000 °/sec (dps), for the accelerometer with ranges of ±2 g, ±4 g, ±8 g and ±16 g, and a single setting of ±1200$\mu T$ range for a compass. For our application the acceleration experienced by the payload does not exceed 2 g as was observed in all of our previous flights, therefore we have chosen the default ±2 g settings for the accelerometer. We have selected the default value of ±2000 *dps* for the gyroscope.

The data from the sensors are digitized by 16-bit analog-to-digital converters (ADCs) for the gyroscope, three 16-bit ADCs for the accelerometer and three 13-bit ADCs for the magnetometer. The data stream from the MPU-9150 is fed into the I$^2$C (Inter-Integrated Circuit) port of the Raspberry Pi. This port is a multi-master serial single-ended computer bus, through which low-speed peripherals are attached to the motherboard using bidirectional open drain lines SDA (Serial Data Line) and SCL (Serial Clock) with pull-up resistors. The DMP application processor performs the sensor fusion by integrating the output from the gyroscope and accelerometer, and generates the quaternions (by InvenSense licensed codes inside the DMP) output which can be read out from the First-In-First-Out (FIFO) registers by our SBC. Since now the sensor timing synchronization and sensor fusion computations are carried out by the DMP, the total load on Raspberry Pi is reduced.

One of the great advantages of the Raspberry Pi is the extensive library of open-source programs available for the platform. We use here the open source code *linux-mpu9150* (developed by the Pansenti software consultancy and development company[3]) to read the data from MPU-9150, combine the magnetometer data with the data from the two other sensors, and calculate the attitude information in terms of Euler angles.

The Euler angles give the orientation of the sensor relative to an Earth-centered inertial frame (ECI), which we convert to celestial coordinates using the position data (latitude and longitude of the payload) obtained from the GPS sensor (see Appendix A). We have used a 20-channel GPS receiver from iWave[4] operating in the L1 frequency band (1575.42 MHz), which gives accuracy better than 10 meters on the ground and 20-30 meters in altitude. The data from the GPS sensor is transmitted to the Raspberry Pi through the Pi's Universal Asynchronous Receiver/Transmitter (UART) port (Fig. 1, *Right*) serially at the rate of 9600 bps. The entire system is powered by a 5 V lithium polymer battery. We have written a program to combine the Euler angles with the GPS values and write the output to a file on a SD card. A detailed description of the algorithm and the program flowchart is presented in Appendix A.

## 3. Attitude Sensor Calibration

### 3.1. *Initial calibration*

The sensor must be calibrated before its first use at a given location by slowly moving it through a range of azimuths and elevations. The program *linux-mpu9150* finds the direction of the magnetic North from the magnetometer values and the local gravitational field vector through the accelerometer output, and then internally defines a local Earth-centered coordinate system. This calibration is sensitive to the local magnetic fields and is affected by the presence of nearby ferric objects, for example, the telescope mount and, therefore, has to be repeated for any change of local conditions.

### 3.2. *Gyroscope calibration*

The attitude sensor can operate in two modes: with and without magnetometer. We initially tested the sensor without the magnetometer. The test involved measuring the drift in elevation and azimuth over a period of 10 hours. A drift of about one arcminute per hour was observed in elevation values which was well within our accuracy limits (see Fig. 2, *Top*). The drift was much higher in azimuth (Fig. 2, *Bottom*) – almost 150 degrees in 10 hours, which is unacceptable for our applications. These drifts are due to two sources of errors — bias [Groves, 2013] and integration errors [Shiau et al., 2012; Borenstein et al., 2008], that arise because of manufacturing limitations of the MEMS gyroscopes [Shiau et al., 2012; Borenstein et al., 2008]. Drifts that are due to the integration errors are linear and can be easily modelled. However, bias errors are due to the intrinsic instability of the gyroscopes and exist even when the gyroscopes are not rotating. In an ideal case we would expect the bias error to be zero, or at least constant, which can be compensated for by subtracting the bias from the output, but in practical situations the bias error varies with each operation of the sensor due to changes in the environment. These errors can be avoided by using higher-precision gyroscopes, such as optical or mechanical ones. However, MEMS gyroscopes have major advantages in having a lower power consumption, a shorter start-up time, and being smaller, lighter and cheaper. Bias errors are usually corrected using other sensors (such as accelerometers) in tandem with MEMS gyroscopes. In situations where accelerometers give a constant value (or zero) (e.g. in yaw angle measurements where there is no acceleration in z-axis), these errors can be avoided by using a fixed external reference — the magnetometer (Fig. 3).

---

[3] https://github.com/richards-tech/linux-mpu9150
[4] iWave Systems SiRF StarIII GSC3f GPS receiver, iWave Systems, India (http://www.iwavesystems.com)

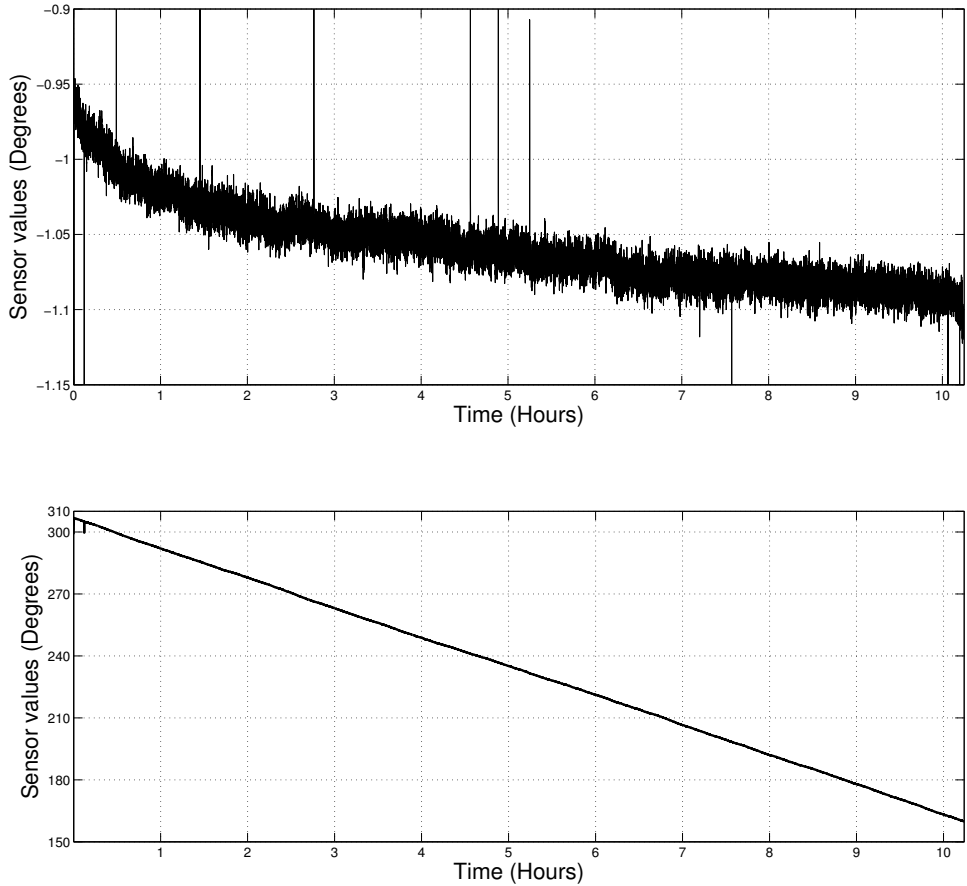

Fig. 2.  Results of the sensor calibration without the magnetometer. *Top*: drift in elevation, *Bottom*: drift in azimuth.

### 3.3. *Magnetometer calibration*

We found the magnetometer in the IMU to have a nonlinear response (Fig. 4) leading to an error of 5 to 10 degrees in azimuth pointing due to cross-field effects and hysteresis [Pang et al., 2013]. To measure the nonlinearity, the sensor was mounted on a 3.5-inch Meade ETX 90 reflector telescope which was rotated by 360 degrees. Sensor values and actual rotation (from the telescope markings) at different positions were noted and used to generate the output. The measured nonlinearity was fitted with a fifth degree polynomial and the corrections were incorporated into our processing algorithm to produce a linear output. This nonlinearity is sensor dependent and must be corrected for each sensor.

### 3.4. *Pointing calibration*

We tested the pointing accuracy of the sensor as part of its calibration. The sensor was mounted on a 3.5-inch Meade ETX 90 reflector telescope and pointed to several known objects in the sky. The same object was observed several times for different elevations. The differences between the actual values and the measured sensor values and the errors are given in Table 2. The average of RMS errors was found to be 0.479 degrees. This test helps us to determine the reproducibility and the accuracy of sensor pointing. The limiting factor in accuracy for both RA and Dec is due to large errors in azimuth values. These errors occur due to the intrinsically poor magnetometer accuracy [Pang et al., 2013], which is ultimately the limiting factor in the overall pointing accuracy of the sensor. A better accuracy may be obtained if *North* is defined independently without the magnetometer, however in this case, the gyroscope introduces a drift due to bias errors, as discussed earlier.

Table 2. Results of the absolute pointing calibration. The values are in DD:MM:SS.

| Object | RA J2014.15 | Dec J2014.15 | Observed RA | Observed Dec | Average RMS Error (Degrees) |
|---|---|---|---|---|---|
| Capella | 79:25:59.88 | 46:00:39 | 78:46:17.15 | 46:05:32.49 | 0.5659 |
|  |  |  | 78:54:14.29 | 45:28:25.52 |  |
|  |  |  | 79:27:41.98 | 46:34:29.10 |  |
| Menkalinan | 90:08:29.98 | 44:56:50 | 89:40:22.04 | 44:52:53.88 | 0.3878 |
|  |  |  | 89:43:48.83 | 44:43:41.29 |  |
|  |  |  | 89:47:13.06 | 44:34:22.42 |  |
| Aldebaran | 69:10:59.99 | 16:30:30.7 | 69:13:29.82 | 16:46:00.34 | 0.5301 |
|  |  |  | 69:32:51.90 | 16:58:48.76 |  |
|  |  |  | 69:47:40.92 | 16:12:49.00 |  |
| Rigel | 78:48:15.02 | -08:11:11 | 78:36:30.17 | -08:15:59.95 | 0.3804 |
|  |  |  | 78:26:08.09 | -07:48:12.64 |  |
|  |  |  | 78:57:53.21 | -07:53:41.54 |  |
| Jupiter | 101:30:29.88 | 23:14:44 | 101:39:16.92 | 22:37:45.79 | 0.5309 |
|  |  |  | 101:38:32.28 | 22:45:58.32 |  |
|  |  |  | 101:16:35.40 | 22:54:22.14 |  |

To test the reproducibility of the sensor output, we have used the readily available Goniometer Stages from Newport$^{TM}$ (with accuracy of 10'), and Thorlabs$^{TM}$ (with accuracy of 1'), to move the sensor in the y-axis (tilt) and z-axis (pan). Sensor readings were taken at some predefined values of azimuth and elevation (pointing). After random movement of the stages, the sensor was brought back to the initial pointing and the readings were recorded. This step was performed five times in total for several different directions. The results are summarized in Table 3.

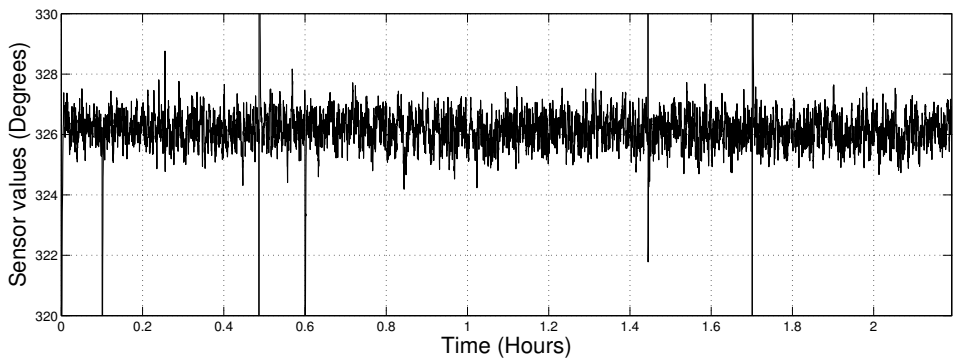

Fig. 3. Results of the sensor calibration with the magnetometer. There is now no drift in the azimuth.

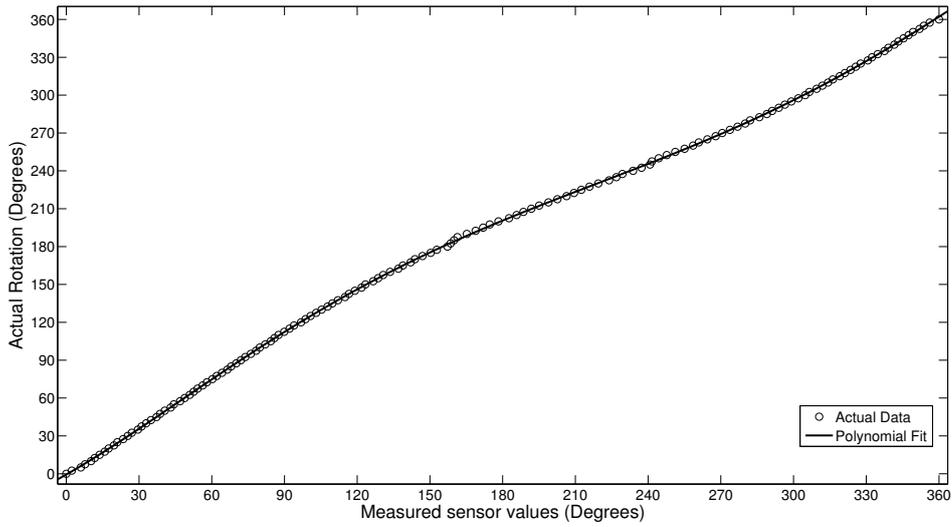

Fig. 4.  Magnetometer nonlinearity.

Table 3.  Results of relative pointing. The given values are in Degrees.

|  | Azimuth | Elevation | Average Azimuth | Average Elevation | Standard deviation in Azimuth | Standard deviation in Elevation |
|---|---|---|---|---|---|---|
| Set 1 | 252.5326 | 1.1952 |  |  |  |  |
| Set 2 | 252.1189 | 1.1983 |  |  |  |  |
| Set 3 | 252.2308 | 1.1706 | 252.3917 | 1.1868 | 0.23904 | 0.0108 |
| Set 4 | 252.7191 | 1.1876 |  |  |  |  |
| Set 5 | 252.3570 | 1.1853 |  |  |  |  |
| Set 1 | 307.5729 | -09.9565 |  |  |  |  |
| Set 2 | 307.3201 | -09.9624 |  |  |  |  |
| Set 3 | 307.4700 | -09.9614 | 307.3638 | -09.9612 | 0.2077 | 0.0042 |
| Set 4 | 307.4264 | -09.9593 |  |  |  |  |
| Set 5 | 307.0294 | -09.9674 |  |  |  |  |
| Set 1 | 29.9928 | 09.9897 |  |  |  |  |
| Set 2 | 30.0042 | 09.9604 |  |  |  |  |
| Set 3 | 30.3929 | 09.9699 | 30.0712 | 09.9753 | 0.2523 | 0.0109 |
| Set 4 | 30.4891 | 09.9769 |  |  |  |  |
| Set 5 | 29.7343 | 09.9793 |  |  |  |  |
| Set 1 | 347.1065 | 06.6215 |  |  |  |  |
| Set 2 | 346.4374 | 06.6384 |  |  |  |  |
| Set 3 | 346.6514 | 06.6102 | 346.6744 | 6.61822 | 0.2581 | 0.0123 |
| Set 4 | 346.5242 | 06.6096 |  |  |  |  |
| Set 5 | 346.6524 | 06.6114 |  |  |  |  |

It was found that the standard deviation of sensor values in elevation (due to accelerometer and gyroscope) and azimuth (due to magnetometer) were around 0.01° and 0.24°, respectively. Since the accuracy of the Goniometer Stage Newport$^{\text{TM}}$ is 10′, we assume this value to be our overall accuracy in elevation measurement (which is within the absolute pointing accuracy, see Table 2). The accuracy of GPS is around $10''$ and introduces much less error to the attitude sensor.

### 3.5. *Thermal Dependence of the sensor*

According to the manufacturer's specifications, the operational range of MPU-9150 is between $-40°C$ and $+60°C$. Since the ambient temperature in stratosphere can reach $-60°C$, the effect of temperature on other electronics components, such as the GPS module and the SBC have to be tested. From one of our high-altitude balloon flights (see Sec. 4), it was found that the GPS module fails to operate below $-40°C$. In addition, it is well known that the performance of lithium polymer batteries reduces at low temperatures, with $-20°C$ being the limit of a good performance (e.g., Ji et al. [2013]).

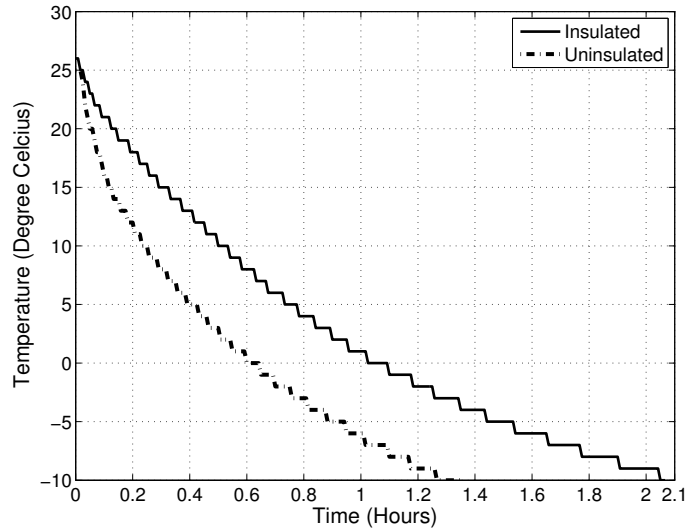

Fig. 5. Temperature test. Solid line — temperature variation in an insulated box. Dashed line — temperature variation in the uninsulated sensor.

We have kept the sensor in an insulated box in a thermal chamber ($-18°C$) for more than two hours. The temperature in the box was maintained above $-10°C$ for 2 hours. When sensor was placed in the chamber uninsulated, the recorded temperature dropped to $-10°C$ in almost one hour (Fig. 5, dashed line). This indicates that we have to take care to properly insulate the equipment in our experiments. From our flight experience, we have found it necessary to balance the internal heating from the electronics with the heat loss from the payload and have experimented with insulated boxes with holes of different diameters to ensures that the temperature are within the allowable ranges.

## 4. Flight Results

This attitude sensor was flown on several high-altitude balloon flights from our launch facility at CREST campus of Indian Institute of Astrophysics (13.1131 N, 77.8113 E). The aim of these experiments was to test the pointing accuracy of a commercial aerial photography mount, intended for astronomical observations from balloons. In Figs. 6 and 7 we present the output of the sensor for the flights conducted on October 13, 2013 and February 16, 2014, respectively. The weight of the payload on the October flight (comprising aerial photography mount with the IMU, attitude sensor, temperature sensors, camera, radio GPS tracker and GSM-GPS tracker) together with the parachute was 5400 gms, and it was carried by two 3000 gms sounding balloons. The balloons were expected to rise above 25 kms at a speed of around 5 m/s, according to manufacture specifications. The balloons rose to a height of about 22 kms in 2 hours after which we beleive one balloon has burst. The remaining balloon attained a state of neutral buoyancy, causing it to float at about that altitude for more than 12 hours, carrying the payload along, as the cut-off mechanism to detach the parachute from the balloons failed. The payload was tracked by radio till it crossed over to the Arabian sea, around 400 kms west of the launch site. The payload was later partially recovered by Indian fishermen from the Arabian sea. While the aerial photography mount with its data was lost, the attitude sensor data was recovered and analyzed. We present the results of the analysis in Fig. 6. Oscillations in azimuth values (Fig. 6, *Top*) indicate that the payload experienced spinning motion during the entire flight with a frequency of 50 degrees in 2–3 minutes, and the elevation values (Fig. 6, *Bottom*) indicate that the payload was experiencing swinging motion with frequency below 2 degrees for the entire flight.

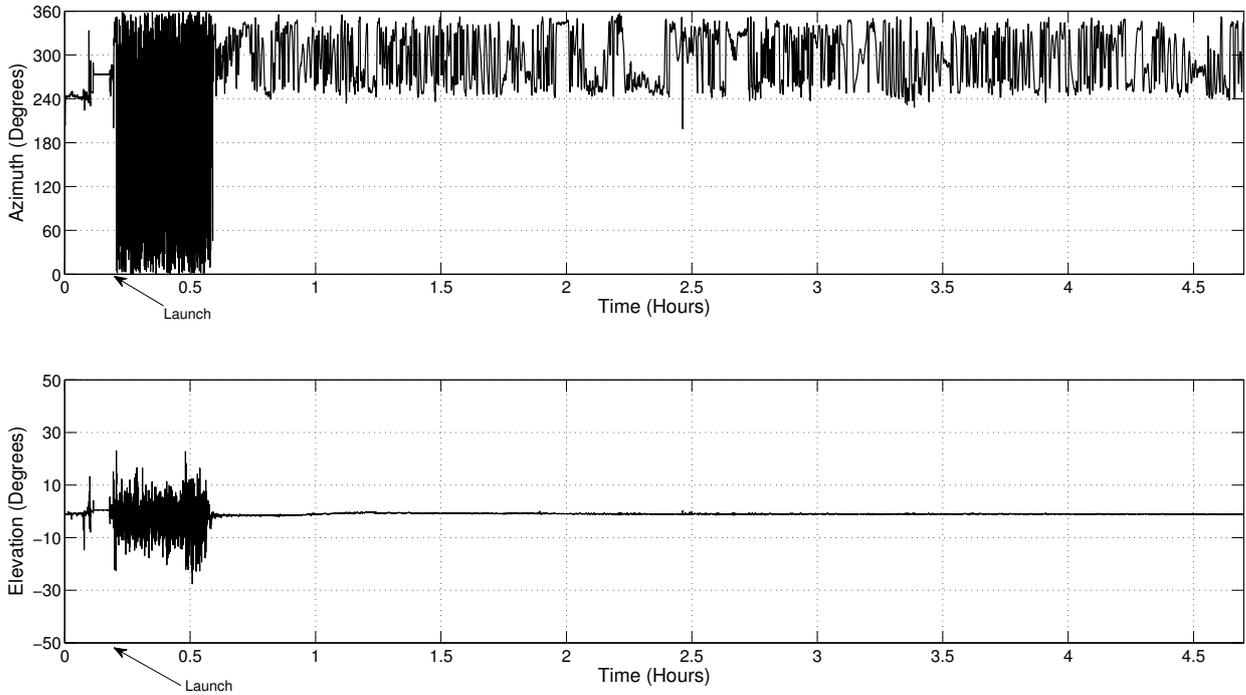

Fig. 6. Attitude sensor data from balloon flight on October 13, 2013. *X*-axis is time of the balloon flight in hours. *Top*: Azimuth values, *Bottom*: Elevation values. The sensor was switched on at 0.0 hours and the balloon was released at 0.2 hours. The large variation in both azimuth and elevation values till 0.2 hours are due to the handling of the payload on the ground.

The payload on the February flight (UV spectrograph, attitude sensor, temperature sensors, camera, radio-GPS tracker and GSM-GPS tracker) together with the parachute weighed 5400 gms and was carried by three 2000 gms sounding balloons. The balloons reached the maximum height of 26.9 kms in 2.5 hours. The balloons burst at that altitude and the payload was successfully recovered around 100 km from the launch site. There was much more turbulence on the second flight (Fig. 7) mainly because of the use of three balloons instead of two and a stormy weather. The change in frequency of swinging and swaying of the payload as it crossed the tropospheric boundary at around 1.2 hours from the start is clearly visible in Fig. 7.

The attitude sensor was flown on these flights in order to test its working in flight conditions and to determine the movement of the payload due to winds. A stabilization platform for astronomical observations is currently in development by our group and eventually the attitude sensor will be flown with it to test its pointing accuracy.

## 5.  Conclusions and Future Work

We have described the development and calibration of a low-cost attitude sensor which can be built with readily available commercial components. The attitude sensor described in this paper can be used as a building block in a closed-loop pointing and stabilization platform for balloon-borne payloads. The cost of our equipment is 5 to 10 times less compared to commercially available products that have similar characteristics and performance[5].

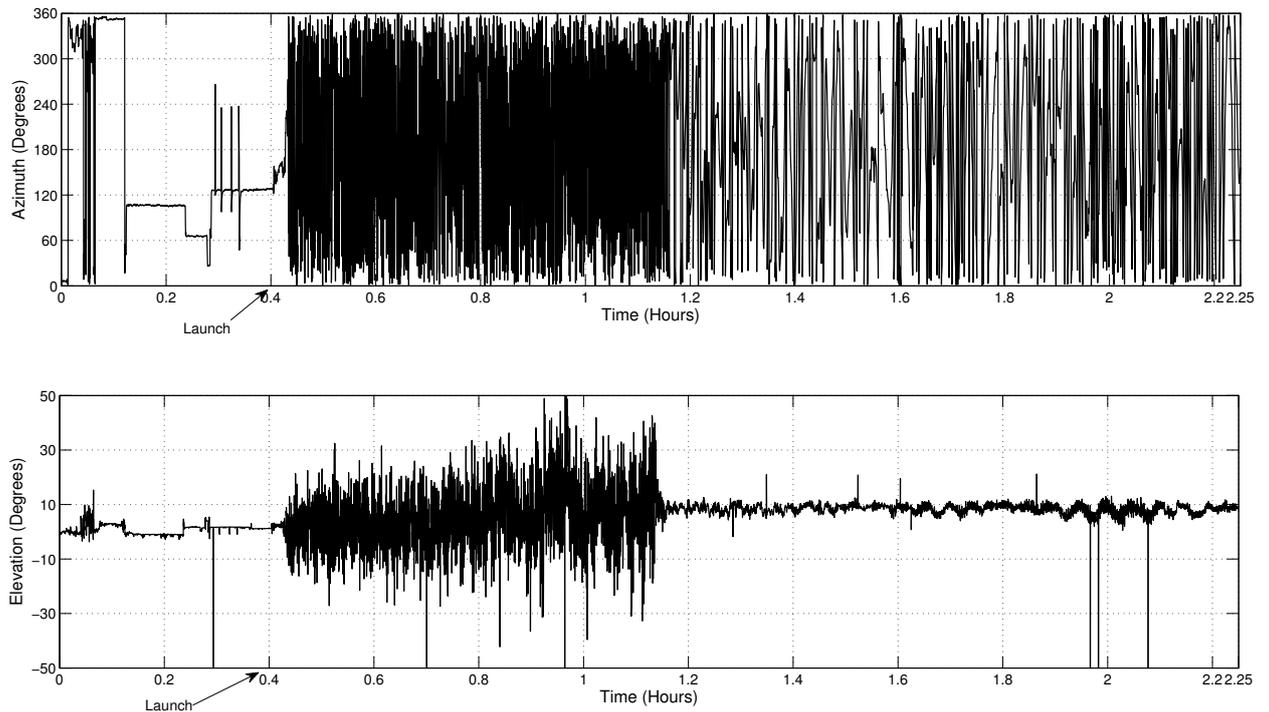

Fig. 7. Attitude sensor data from balloon flight February 16, 2014. X-axis is time of the balloon flight in hours. *Top*: Azimuth values, *Bottom*: Elevation values. The sensor was switched on at 0.0 hours and the balloon was released at 0.4 hours.

The overall accuracy of the developed sensor was found to be 0.5° due to the intrinsic errors in magnetometer. We plan to improve the accuracy by operating the attitude sensor without the magnetometer, defining the initial *North* direction prior to the launch by other means and correcting for the gyroscope errors (Sec. 3.2) in the algorithm. The accuracy of the system can also be improved by using high-precision, but more expensive, magnetometers, accelerometers and speed sensors [Gebre-Egziabher et al., 2004]. However, the current sensor can be used for initial coarse pointing on a stabilized platform. Work for such a pointing system with stepper motors and/or reaction wheels is currently in progress.

The sensor fusion employed in the algorithm has been taken from already available open-source codes. It will be worthwhile to develop data filtering and data fusion algorithms of our own, which may improve the accuracy of the sensor. The experience gained in the development and testing of this device will further enhance our ability in the development of similar, yet more complex, systems. Our group is also developing a low-cost star tracker, where this attitude sensor will be used for the initial pointing. The details of these works will be discussed in the forthcoming paper.

**Acknowledgments**

We thank the anonymous referees for very helpful comments and suggestions that considerably improved the paper.

---

[5]e.g., VN-200 *http://www.vectornav.com/purchase/volume-pricing*

## Appendices

## Appendix A  Description of the software

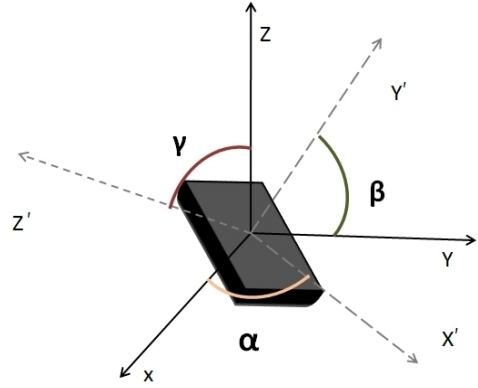

Fig. A.1.  Attitude sensor reference frame. The axes $x$, $y$ and $z$ define the ECI reference frame. The axes $x'$, $y'$ and $z'$ define body-centered reference frame. Angles $\alpha$, $\beta$ and $\gamma$ are the Euler angles.

We have used in our attitude sensor the software package *linux-mpu9150* developed by the Pansenti software company for the Raspberry Pi under Raspbian Linux platform. The code reads the data from the MPU-9150 and calculates the Euler angles. Here, the data fusion from accelerometer and gyroscope is done onboard the MPU-9150 while the magnetometer data fusion is done in the main code. The code operates by opening the I$^2$C bus for communication between the Raspberry Pi and the MPU-9150. It then reads the data from the IMU and uses the Euler angle transformation [Pio, 1966] to calculate the yaw $\alpha$, pitch $\beta$ and roll $\gamma$ angles (Fig A.1). The roll and pitch angles are derived from the accelerometer and gyroscope output, and the yaw readings from the magnetometer. The accelerometer + gyroscope data are saved in the form of quaternions, which are then read from the DMP FIFO registers to calculate Euler angles using the following formulae

$$pitch(\beta) = \arcsin(2 \times (ac - bd)), \tag{A.1}$$

$$roll(\gamma) = \arctan\left\{\frac{2 \times (cd + ab)}{1 - 2 \times (b^2 + c^2)}\right\}, \tag{A.2}$$

where $a, b, c, d$ are the quaternions. After that, the magnetometer output is used to calculate the yaw angle,

$$yaw(\alpha) = \arctan\left\{\frac{MagY}{MagX}\right\}, \tag{A.3}$$

where MagY and MagX are the Y-axis and X-axis magnetometer readings, respectively. We have used ±2000 dps and ±2 g settings for gyroscope and accelerometer, respectively. We have modified package *linux-mpu9150* to calculate the equatorial coordinates (RA and Dec) of the pointing by combining the IMU and the GPS data (the source code is available at the following address (https://github.com/iiabaloongroup). Our code calculates the elevation *Alt* and azimuth *Az* of the horizontal coordinate system, which defines the absolute pointing of the sensor with respect to an Earth-centered coordinate system, from the Euler angles. The latitude $\phi$ and longitude $l$ of the payload and the current universal time (UTC) are then read from the GPS module using the National Marine Electronics Association (NMEA) protocol over the UART port. The local sidereal time (LST) of observation in degrees is then calculated from the UTC by

$$\text{LST} = \text{GMST} + \text{UTC} \cdot 15.0 + l, \tag{A.4}$$

where GMST (Greenwich Mean Sidereal Time) in degrees is

$$289.9404 + 4.70935 \times 10^5 \cdot d + 356.0470 + 0.9856002585 \cdot d + 180 \,, \tag{A.5}$$

and quantity $d$ is calculated from the date information YYYY-MM-DD as

$$d = 367 \cdot \text{YYYY} - \frac{7}{4}\left(YYYY + \frac{MM+9}{12}\right) + \frac{275 \cdot MM}{9} + DD - 730530.0 + \frac{\text{UTC}}{24}\,. \tag{A.6}$$

After LST is calculated, the equatorial coordinates RA and Dec are obtained from the following formulae

$$Dec = \arcsin\{\sin(Alt)\sin\phi + \cos\phi\cos h\cos(Az)\}\,, \tag{A.7}$$

$$HA = \arccos\{\cos\phi\sec(Dec) - \tan(Dec)\tan\phi\}\,, \tag{A.8}$$

$$RA = LST - HA\,, \tag{A.9}$$

where HA is the hour angle. The output of the attitude sensor contains the following information: equatorial coordinates (RA and Dec) of the source, current time in UTC, GPS coordinates (latitude and longitude) of the payload, azimuth and elevation of the pointing direction. The flowchart of the complete algorithm is presented in Fig A.2.

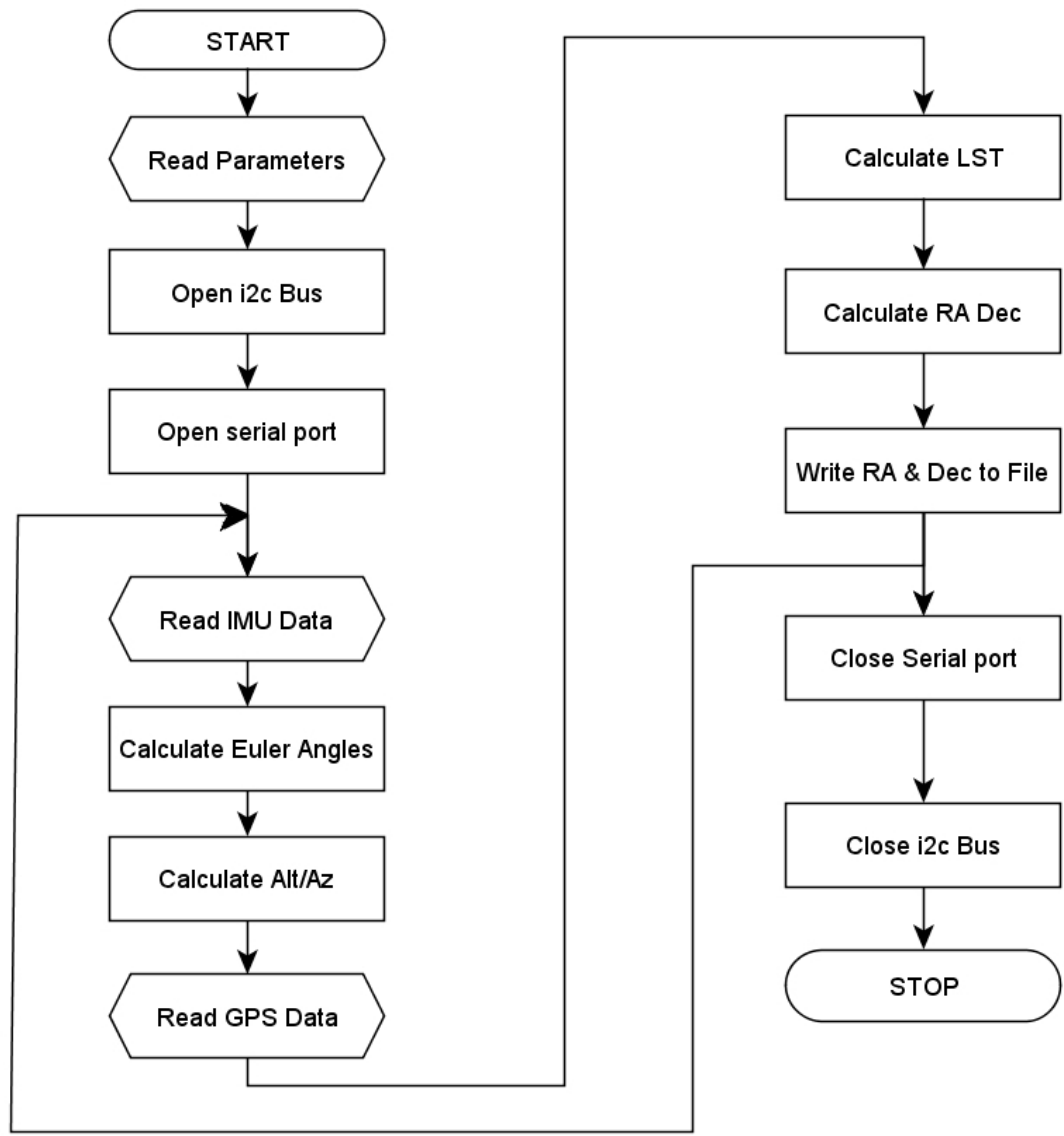

Fig. A.2.  Flowchart of the attitude sensor code.